\documentclass[aps,pra,onecolumn,preprint,showpacs,groupedaddress,superscriptaddress,nofootinbib,floatfix,preprintnumbers,longbibliography]{revtex4-2}

\usepackage{physics}
\usepackage{xcolor}
\usepackage{graphicx}
\usepackage{subfigure} 
\usepackage{caption}
\usepackage{dcolumn}
\usepackage{bm}
\usepackage{multirow}
\usepackage{amssymb}

\usepackage{comment}
\usepackage{ulem}

\begin{document}

\title{Free temporal evolution of a \textit{superbandwidth} wave packet: The quantum tale of the hare and the tortoise 
 }

\author{Enrique G. Neyra}
\email{enriqueneyra@cnea.gob.ar}
   \affiliation{Instituto Balseiro (Universidad Nacional de Cuyo and Comisión Nacional de Energía Atómica) and CONICET CCT Patagonia Norte. Av. Bustillo 9500, Bariloche 8400 (RN), Argentina.}

\author{Marcelo F. Ciappina}
\email{marcelo.ciappina@gtiit.edu.cn}
\affiliation{Department of Physics, Guangdong Technion - Israel Institute of Technology, 241 Daxue Road, Shantou, Guangdong, China, 515063}
\affiliation{Technion -- Israel Institute of Technology, Haifa, 32000, Israel}
\affiliation{Guangdong Provincial Key Laboratory of Materials and Technologies for Energy Conversion, Guangdong Technion - Israel Institute of Technology, 241 Daxue Road, Shantou, Guangdong, China, 515063}

\author{Lorena Reb\'on}
\email{rebon@fisica.unlp.edu.ar}
\affiliation{ 
Departamento de F\'isica, FCE, Universidad Nacional de La Plata, C.C. 67, 1900 La Plata, Argentina
}%
\affiliation{Instituto de F\'isica de La Plata, UNLP - CONICET, Argentina}

\date{\today}

\begin{abstract}

Can the interplay between quantum mechanics and classical optics offer new perspectives on wavepacket dynamics? Building on this connection, we show that local momenta with both superoscillatory and suboscillatory characteristics can arise during the free propagation of a quantum particle. This behavior is mathematically analogous to the propagation of a superbandwidth laser pulse in a dispersive medium, where the instantaneous frequencies of the electric field exhibit similar sub- and superoscillatory features. This analogy is rooted in the equivalence between the wave description of light in classical optics and the formalism of quantum mechanics. In addition, we explore the phenomenon of wavepacket localization during free propagation within a bounded region. This localization is directly linked to the distribution of local momenta within the confined wavepacket region. To complement our quantum mechanical analysis, we also perform a classical analysis to provide further insights into this phenomenon. Our findings reveal that both the emergence of local momenta with sub- and superoscillatory features and the wavepacket localization occur within a distinct timescale, which we define as the \textit{interference time}.

\end{abstract}

\maketitle

\section{Introduction}

Understanding how waves interact with their environment is a cornerstone of physics,
underpinning developments in fields ranging from telecommunications to quantum
mechanics. Waves are characterized by their spectral content, which determines
properties like frequency and bandwidth. In conventional scenarios, the behavior of a
wave is tightly linked to its Fourier spectrum, a mathematical representation of its
frequencies. However, recent discoveries have begun to challenge this classical
paradigm, revealing phenomena where waves locally exhibit characteristics that seem
to defy their spectral limits ~\cite{chen2019superoscillation,zheludev2022optical,berry2019roadmap}. 

The connection between wave phenomena and quantum mechanics is remarkably profound. In quantum mechanics, particles such as electrons and photons are described by wavefunctions, which dictate their probabilistic behavior and evolution. These wavefunctions follow the same mathematical principles as classical waves: solutions to the Schrödinger equation can be analogized to the propagation of light in dispersive media.
This shared foundation allows insights from one domain to inform the other, and effects
like superposition, interference, and dispersion emerge naturally in both. For instance,
concepts like group velocity dispersion in optics have direct analogs in quantum
mechanics, where the spread of a particle’s wave packet reflects similar principles.
These parallels enable us to explore wave behavior through a unified lens, bridging
classical and quantum systems~\cite{Forbes2019}.

One such concept, recently introduced in the field of classical optics, is \textit{superbandwidth}. This phenomenon describes how a band-limited signal—a signal confined to a specific frequency range—can locally behave as if its bandwidth is broader than its Fourier spectrum indicates~\cite{Neyra_21}. For instance, sub-Fourier laser pulses with superbandwidth characteristics have been shown to interact with matter as though they encompass a continuum of frequencies outside their defined spectral range~\cite{Neyra_22}, extending both above and below the boundaries of the Fourier spectrum. This surprising effect arises from destructive interference among pulses with differing temporal and amplitude profiles, a behavior observed not only in Gaussian pulses but also in more complex non-Gaussian shapes.

The implications of this phenomenon become even more intriguing when superbandwidth pulses propagate through dispersive media, where wave packets spread over time due to frequency-dependent velocities. A recent study~\cite{Neyra2023} demonstrated that, in such scenarios, superbandwidth pulses generate localized oscillations of the electric field, with instantaneous frequencies that exceed or fall below their spectral limits. Moreover, by fine-tuning pulse synthesis parameters, these oscillations can emulate a reversal of the medium’s dispersion properties—causing higher frequencies to travel faster than lower ones, or vice versa, regardless of the medium’s inherent characteristics.

These findings not only challenge conventional understandings of wave dynamics but also resonate with analogous phenomena in quantum mechanics. Examples include \textit{superoscillations}, which are well-documented in the literature~\cite{aharonov2017mathematics,berry2006evolution,yuan2016quantum,bloch2023spacetime}, and the \textit{backflow} effect~\cite{Berry_2010,yearsley2013introduction,penz2005new,zhang2025observation}, which involves the counterintuitive reversal of probability flow in a quantum system. Both phenomena underscore the surprising consequences of wave superposition and have parallels in classical optics~\cite{Eliezer2017, chen2019superoscillation,zheludev2022optical,gbur2019using,wen2014super, Eliezer_20, Daniel_2022}.

In this work, we explore the quantum analog of the superbandwidth effect, leveraging the significant connections between the mathematical frameworks governing waves in optics and quantum mechanics. Specifically, we examine the propagation of quantum wave packets and identify phenomena that parallel the dynamics of superbandwidth pulses in dispersive media. By bridging these two domains, our study provides new insights into the intricate interplay between spectral content and spatial dynamics in wave-like systems, paving the way for innovative applications in quantum and optical technologies.

\section{Free particle evolution and local momentum}

In what follows, we analyze the propagation of a non-relativistic particle of mass $m$, whose quantum state is described by the wavefunction $\psi(x,t)$ as it moves freely along the $x$-axis. We assume that, at $t = 0$, the wavefunction $\psi(x,0) \equiv \psi_0(x)$ corresponds to a function that exhibits superbandwidth~\cite{Neyra_21}. To this end, we begin by defining the initial momentum distribution of the state:
\begin{equation}
 \phi_0(\kappa)  = \mathcal{N}^{-1}\left(e^{-(\frac{\kappa-\kappa_0}{\Delta\kappa})^2} - \alpha e^{-(\frac{\kappa-\kappa_0}{\Delta\kappa/2})^2}\right),
\label{eq:wave_packet_momentum}
\end{equation}
where $\kappa$ represents the wavenumber, which is related to the momentum of the particle ($p$) through the de Broglie relation~\cite{deBroglie}: $\kappa = \frac{p}{\hbar}$. The normalization constant $\mathcal{N}$ is expressed as: $\mathcal{N} = \left(\frac{\pi}{4}\right)^{1/4} \sqrt{\Delta\kappa\left( \sqrt{2} - \frac{4\alpha}{\sqrt{5}} + \frac{\alpha^2}{2}\right)
}$. 

As shown, a quantum state described by the momentum representation in Eq.~\eqref{eq:wave_packet_momentum} is a superposition of two Gaussian wave packets with central momentum $p_0 = \hbar \kappa_0$, where $\Delta\kappa$ is proportional to their standard deviation. The position representation of the initial state is then given by:
\begin{equation}
 \psi_0(x) = \mathcal{F}\{\phi_0(\kappa)\} = \mathcal{N}^{-1}\frac{\Delta\kappa}{\sqrt{2}} \left(e^{i\kappa_0 x}e^{-(\frac{\Delta\kappa x}{2})^2} - \frac{\alpha}{2}  e^{i\kappa_0 x}e^{-(\frac{\Delta\kappa x}{4})^2}\right),
\label{e2}
\end{equation}
where $\mathcal{F}\{\cdot\}$ indicates the Fourier transform. In free-space, this state evolves according to the one-dimensional time-dependent Schr\"odinger equation (1D-TDSE) 
\begin{equation}
 i\hbar\frac{\partial \psi(x,t)}{\partial t}= -\frac{\hbar^2}{2m}\frac{\partial^2 \psi(x,t)}{\partial x^2}~,
\label{eq_Schrodinger}
\end{equation}
whose solution can be obtained by means of the free-particle propagator $G(x,t;x',t')$ (Green's function) as:  
\begin{eqnarray}
 \psi(x,t)&=&\int_{-\infty}^{\infty} dx' ~ G(x,t;x',0) \psi_0(x') \nonumber\\
 &=&\frac{m}{\sqrt{2\pi i \hbar t}}\int_{-\infty}^{\infty} dx' ~ e^{i\frac{m}{2\hbar t}(x-x')^2} \psi_0(x') ~,
\label{e4}
\end{eqnarray}
or equivalently,
\begin{eqnarray}
 \psi(x,t) = \mathcal{F}\{e^{-i\frac{\hbar k^2}{2 m }t}\phi_0(k)\}~.  
 \label{eq:e5}
 \end{eqnarray}
Then, Eq.~\eqref{eq:e5} describes the temporal evolution of two wavepackets that exhibit destructive interference (see Eq.~\eqref{e2}). Both are centered at $\kappa_0$, with standard deviations of $\Delta\kappa$ and $\Delta\kappa/2$, respectively. In the position representation, the information about the particle's momentum is lost. However, $\psi(x,t)$ can be associated with a local momentum or, alternatively, with the local wavenumber $\kappa_l(x,t)$, defined as~\cite{Berry_2013}:
\begin{equation}
 \kappa_l(x,t)= \mathrm{Im} \frac{\partial}{\partial x} \mathrm{log}(\psi(x,t))=\frac{\partial}{\partial x} \mathrm{Arg} (\psi(x,t))~.
\label{eq:local_momentum_b}
\end{equation}

\section{Results}\label{suppl_peak}

We present here the main features of the free evolution of a quantum state with a superbandwidth wavefunction, as described by Eq.~\eqref{eq:e5}.
Without loss of generality, the calculations have been performed in units where $\hbar=1$, and the particle mass has been set to $m = 1$.
It is important to note that, under the condition $\hbar=1$, the momentum and the wavenumber coincide in their numerical values, allowing us to refer to them interchangeably.
Additionally, we have set $\kappa_0 = 2\pi$, which implies that the spatial variable $x$ is expressed in units of the de Broglie wavelength $\lambda_0$, with $\lambda_0 = 2\pi/\kappa_0=1$. Consequently, the temporal units in this unit system can be derived from the relation $[t]=[m][\lambda_0^2]/[\hbar]$.

\begin{figure*}[h!]
\centering
\includegraphics[width=0.8\textwidth]{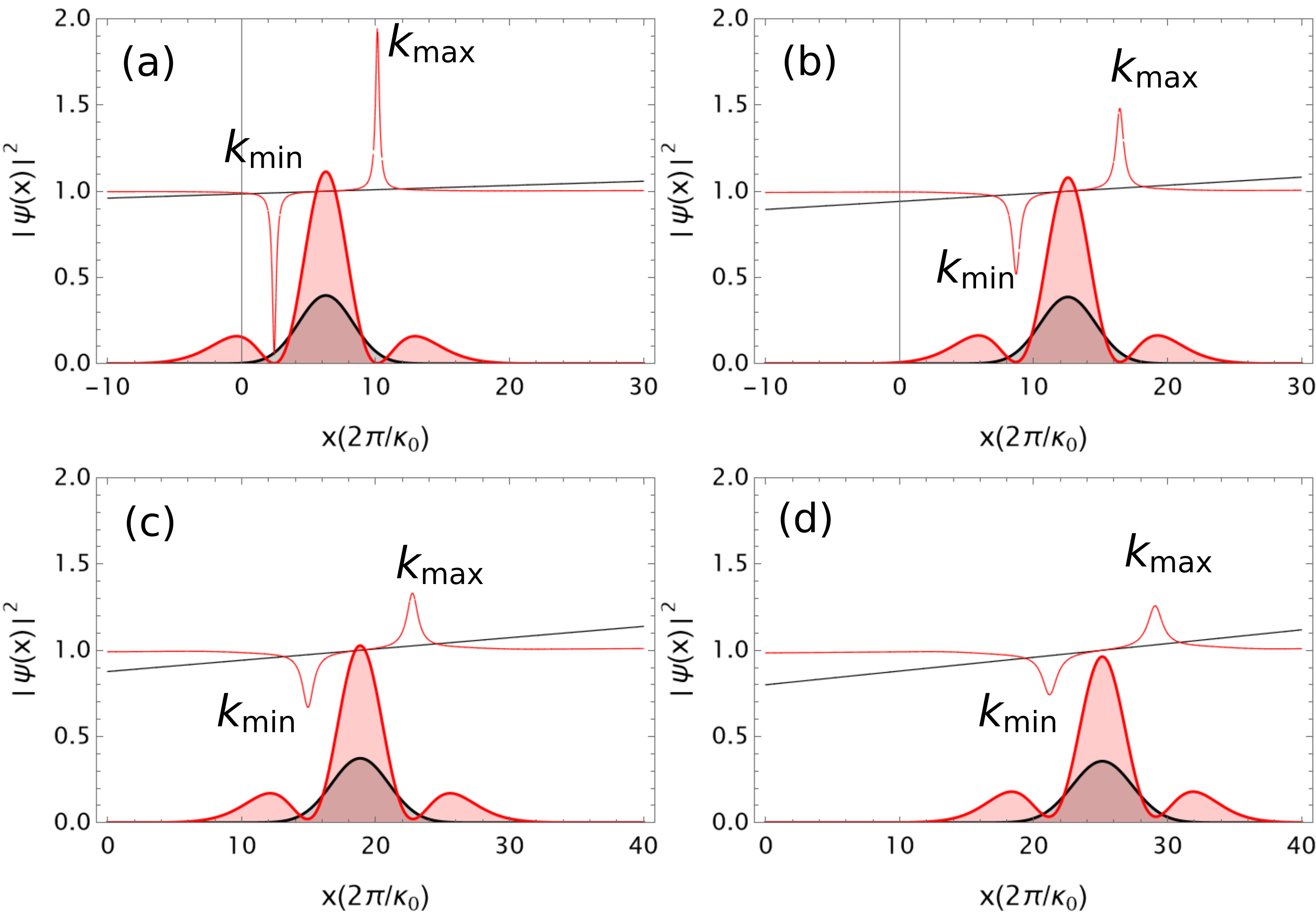}
\caption{Free evolution of a superbandwidth quantum state. Thick red lines indicate the probability density $|\psi(x,t)|^2$ at different fixed times, where the insets (a), (b), (c) and (d) correspond to the times $t=1$, $t=2$, $t=3$ and $t=4$, respectively ($t=0$ corresponds to the case in which there is no propagation). In all cases, we have set $\Delta \kappa=0.5$ and $\alpha=1$. For comparison, we also plot the density probability of a Gaussian wave packet ($\alpha = 0$ in Eq.~\eqref{e2}), which is depicted by the thick black line. Thin red (black) lines indicate the value of the local momentum $\kappa_l(x,t)$, normalized to $\kappa_0$, for the superbandwidth (Gaussian) state. 
} \label{fig:Fig1SM}
\end{figure*}

In Figs.~\ref{fig:Fig1SM}(a) to~\ref{fig:Fig1SM}(d), we present the probability density $|\psi(x,t)|^2$ (thick red line) at different times $t$, with the synthesis parameters of the superbandwidth state set to $\Delta\kappa = 0.5$ and $\alpha = 1$. In each panel, the local momentum $\kappa_l(x,t)$, normalized to $\kappa_0$, is shown as a thin red line. For comparison, the probability density of the Gaussian state (corresponding to $\alpha = 0$ in Eqs.~\eqref{eq:wave_packet_momentum}, \eqref{e2}, and~\eqref{eq:e5}), along with its corresponding local momentum, is depicted by thick and thin black lines, respectively. Notably, two distinct local momenta, labeled as $\kappa_{max}$ and $\kappa_{min}$, emerge in the destructive interference region and are clearly associated with the zeros of the superbandwidth wavefunction. These local momenta exhibit the unique property that their values, $\kappa_{max}$ and $\kappa_{min}$, are higher and lower, respectively, than those in the momentum spectrum of the state as described by Eq.~\eqref{eq:wave_packet_momentum}. To illustrate this, we compare the ``weight" of the momentum component $\kappa_{max}$ (or $\kappa_{min}$) in the momentum distribution $\phi_0(\kappa)$ (Eq.~\eqref{eq:wave_packet_momentum}) with the magnitude of the probability density $|\psi(x,t)|^2$ at the point $(x, t)$, where the local momentum $\kappa_l(x,t)$, normalized to $\kappa_0$, equals $\kappa_{max}$ (or $\kappa_{min}$). For instance, in Fig.~\ref{fig:Fig1SM}(a), where $\kappa_{max} = 1.95$, $t = 1$, and $x \equiv x_{max} \approx 10$, we estimate $\sqrt{|\phi_0(\kappa_{max})|^2 / |\psi(x_{max},t)|^2} \approx 1 \times 10^{-60}$. Furthermore, due to the symmetry of the wavefunction, the value $\kappa_l(x,t) = \kappa_{min}$ (with $\kappa_{min} = 2 - \kappa_{max}$) occurs at a position $x$ that is symmetrically opposite to the location of $\kappa_l(x,t) = \kappa_{max}$. Consequently, the same ratio values are obtained for $\sqrt{|\phi_0(\kappa_{min})|^2 / |\psi(x_{min},t)|^2}$.

It should be noted that the appearance of these high $(\kappa_{max})$ and low $(\kappa_{min})$ local momenta (or wavenumbers) can be interpreted as superoscillations of the quantum mechanical wavefunction and the complementary but less explored phenomenon of suboscillations~\cite{Eliezer_17}. However, we demonstrate here that, for a quantum state initially described by a superbandwidth wavepacket and freely evolving under the Schrödinger equation (Eq.~\eqref{eq_Schrodinger}), these sub- and superoscillations represent facets of a single phenomenon, as they occur within the same wavefunction at the same propagation time.
As time progresses, this superoscillatory (or suboscillatory) behavior diminishes. For instance, in the case shown in Fig.~\ref{fig:Fig1SM}(d), where $t = 4$, the value $\kappa_l(x,t) = \kappa_{max} = 1.26$ occurs at $x \equiv x_{max} \approx 29$, with $\sqrt{|\phi_0(\kappa_{max})|^2 / |\psi(x_{max},t)|^2} \approx 7 \times 10^{-4}$. 
\begin{figure*}[h!]
\includegraphics[width=0.8\textwidth]{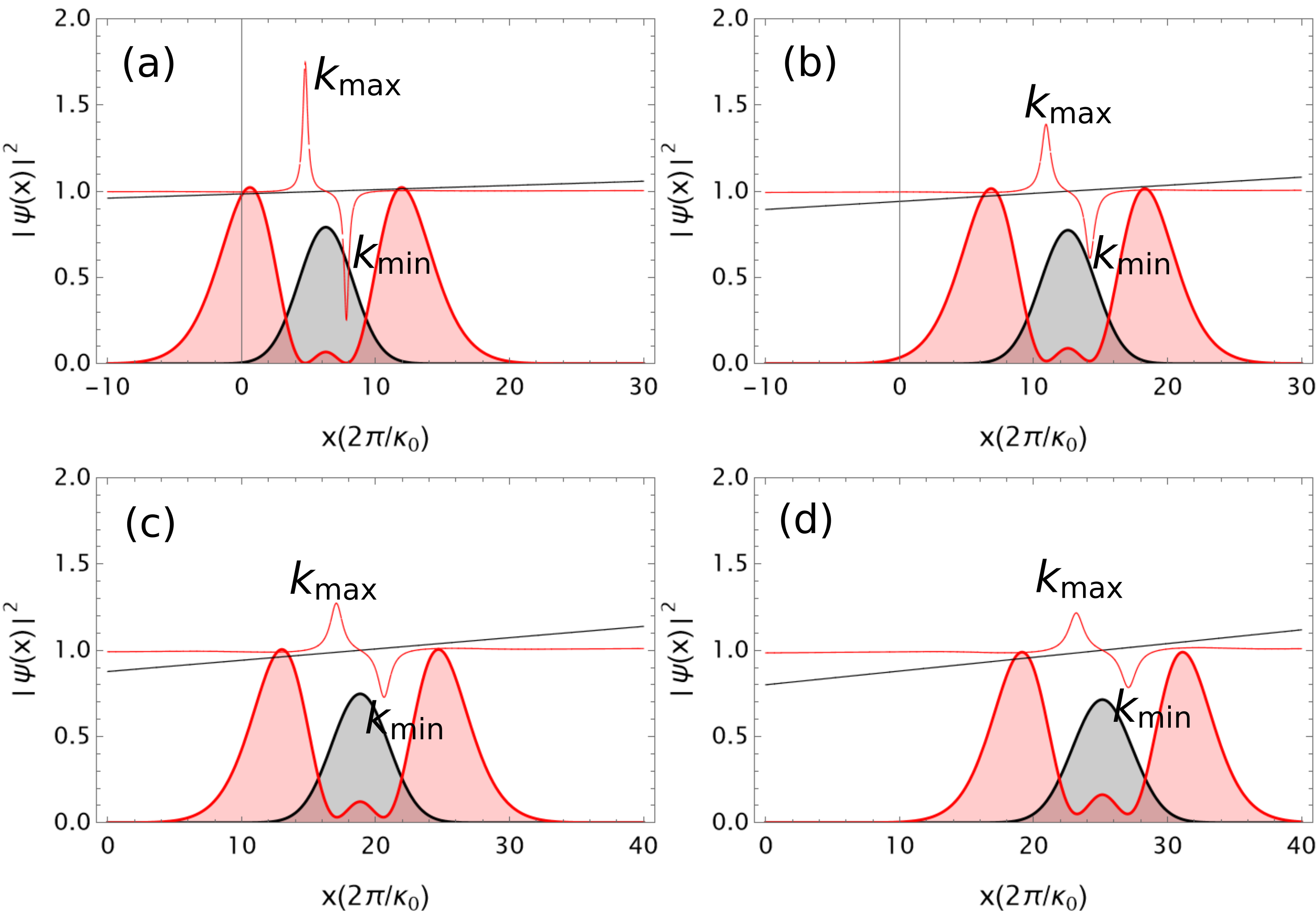}
\caption{Idem as Fig.~\ref{fig:Fig1SM} but for a superbandwidth quantum state
with $\alpha=1.8$.
} \label{fig:Fig2SM}
\end{figure*}

Similarly, sub- and superoscillations are observed in the wavefunction of the state depicted in Figs.~\ref{fig:Fig2SM}(a)--\ref{fig:Fig2SM}(d), where a different value of the parameter $\alpha$ was chosen ($\alpha = 1.8$). In this case, however, the spatial positions of the particle's local momenta $\kappa_l(x,t) = \kappa_{min}$ and $\kappa_l(x,t) = \kappa_{max}$ are reversed compared to the case shown in Fig.~\ref{fig:Fig1SM}. Specifically, it is found now that $x_{min} > x_{max}$. In Table \ref{tb:results}, we quantitatively summarized these findings, together with the mentioned in the previous paragraph.

\begin{table*}[h!]
\centering
\begin{tabular}{c|c|c|c|c|c|c|c}
                                                   & $\;\;\kappa_{max}\;\;$ & $\;\;\kappa_{min}\;\;$ &$\;\;|\phi_0(\kappa_{max})|^2\;\; $ & $\;\;\;t\;\;\;$ & $\;\;x_{max}\;\;$ & $\;\;x_{min}\;\;$ & $\;\;|\psi(x_{max},t)|^2\;\;$  \\ 
                                                   & & & $\;\;\;\;|\phi_0(\kappa_{min})|^2\;\;\;\;$ &  &  &  & $\;\;\;\;|\psi(x_{min},t)|^2\;\;\;\;$\\ \hline\hline
\multicolumn{1}{c|}{\multirow{4}{*}{$\;\;\;\;\;\alpha=1\;\;\;\;$}}   &  1.95   & 0.95   & $\;\;\;2.9\times10^{-123}\;\;\;$ & 1 & $\;\;10.0\;\;$ & $\;\;2.4\;\;$ &$\;\;\;2.0\times10^{-3}\;\;\;$ \\ 
\multicolumn{1}{c|}{}                              &  1.48   & 0.48   & $\;\;\;4.5\times10^{-31}\;\;\;$ &  2  &  16.4 & 8.7 &   $\;\;\;5.0\times10^{-3}\;\;\;$                \\ 
\multicolumn{1}{c|}{}                              &  1.33   &  0.33   & $\;\;\;2.0\times10^{-14}\;\;\;$ & 3 &  22.7 &  14.9 &    $\;\;\;1.0\times10^{-2}\;\;\;$              \\ 
\multicolumn{1}{c|}{}                              &  1.26   & 0.26   & $\;\;\;9.6\times10^{-9}\;\;\;$  &   4  &  29.0 & 21.2 & $\;\;\;1.8\times10^{-2}\;\;\;$               \\ \hline\hline
\multicolumn{1}{c|}{\multirow{4}{*}{$\;\;\;\;\alpha=1.8\;\;\;\;$}} & 1.75 &  0.75 &$\;\;\;1.3\times10^{-76}\;\;\;$ &   1  &  4.7 &  7.8 & $\;\;\;3.1\times10^{-1}\;\;\;$                  \\ 
\multicolumn{1}{c|}{}                              & 1.39 &  0.39 &$\;\;\;2.0\times10^{-20}\;\;\;$  &  2   &  10.9 & 14.2 & $\;\;\;2.9\times10^{-1}\;\;\;$                \\ 
\multicolumn{1}{c|}{}                              & 1.27 & 0.27 & $\;\;\;1.0\times10^{-9}\;\;\;$  &  3   &  17.0 & 20.6 & $\;\;\;2.4\times10^{-1}\;\;\;$                \\ 
\multicolumn{1}{c|}{}                              & 1.21 & 0.21 &   $\;\;\;1.0\times10^{-5}\;\;\;$      &  4   &  23.2 &  27.0 &  $\;\;\;2.2\times10^{-1}\;\;\;$            \\ \hline
\end{tabular}
 \caption{Summary of numerical values of interest associated with the local momentum $\kappa_l(x,t)$, corresponding to the superoscillation ($\kappa_{max}$) and suboscillation ($\kappa_{min}$) of the wavefunction $\psi(x,t)$.}\label{tb:results}
\end{table*}

The results analyzed here are mathematically equivalent to those presented in Ref.~\cite{Neyra2023} within the context of classical optics. In a wave description, the local wavenumber determines the local oscillation frequency, making the analysis discussed above entirely analogous to the optical case by replacing the quantum wavefunction $\psi(x,t)$ with the electric field of an optical pulse, $E(t)$, and $\kappa_l(x,t)$ with its instantaneous frequency, $\omega(t)$. In that work, it was shown that, in the vicinity of the central lobe of the electric field (or equivalently, the quantum mechanical wavefunction), where the local wavenumber can be approximated as a linear function, there exists a critical value of $\alpha$ ($\alpha_C$). For this critical value, the sign of the slope of the linear function changes, leading to a reversal in the relative positions of $\kappa_{max}$ and $\kappa_{min}$. Similarly, we find here that there is a class of quantum states with $\alpha < \alpha_C$, where the dynamics of the suboscillations ($\kappa_{min}$) and superoscillations ($\kappa_{max}$) behave as observed in Fig.~\ref{fig:Fig1SM}. Conversely, for states with $\alpha > \alpha_C$, the opposite behavior is observed, as shown in Fig.~\ref{fig:Fig2SM}.

One of the intriguing questions is how to interpret the paradoxical scenario where the lower local momentum ($\kappa_{min}$) ``seems to travel" faster than the higher local momentum ($\kappa_{max}$), as $\kappa_{min}$ consistently appears ahead of $\kappa_{max}$ at each time $t$. To uncover the underlying physics of this phenomenon, which arises from quantum mechanical wavefunctions with superbandwidth, we analyze it in three sections. First, using the quantum-mechanical current and probability flux concepts, we calculate the probability flux at the positions corresponding to $\kappa_{min}$ and $\kappa_{max}$ and interpret the results in comparison with the behavior of Gaussian wavepackets. This provides insight into the probability density dynamics associated with superbandwidth states. Second, we perform a classical analysis and study the free propagation of an ensemble of classical particles to gain a conceptual understanding of the phenomenon. Here, we also define the \textit{interference time}, which estimates when the local oscillations at $\kappa_{min}$ and $\kappa_{max}$ disappear. This analysis serves as a complementary perspective rather than a semiclassical interpretation. Finally, we investigate the wavepacket evolution through Bohmian mechanics~\cite{Bohm1952}, an alternative quantum framework. This approach describes particle trajectories via a guiding equation influenced by a quantum potential, introducing non-local effects and offering a deterministic perspective distinct from classical mechanics



\subsection{Quantum-mechanical current and probability flux}

In this section, we will calculate the probability flux between the times where the local momenta, centered at the position of $\kappa_{min}$ and $\kappa_{max}$, exist, and interpret these results by comparing them with the behavior of the Gaussian wavepacket.

Since in the quantum mechanical framework $|\psi(x,t)|^2$ represents the probability density, then the probability flux, $F(t_i,t_f)$, that crosses a transverse plane at the position $x_P$ in the time interval $[t_i,t_f]$, can be expressed as
\begin{eqnarray}
 F(t_i,t_f) &=& \int_{-\infty}^{x_P} dx \;|\psi(x,t_i)|^2 - \int_{-\infty}^{x_P} dx\; |\psi(x,t_f)|^2\\
 &=&\int_{t_i}^{t_f} dt\;J(x_P,t)\;,\nonumber
\label{eq:flux}
\end{eqnarray}
where $J(x,t)$ is the quantum-mechanical (probability) current at position $x$, with the usual definition 
\begin{eqnarray}
 J(x,t)&=&\frac{\hbar}{m}\mathrm{Im} \left(\psi^*(x,t)\;\frac{\partial \psi(x,t)}{\partial x} \right)\;,
\label{eq:current}
\end{eqnarray}
which meets the continuity equation:

\begin{eqnarray}
\frac{\partial |\psi(x,t)|^2}{\partial t} + \frac{\partial J(x,t)}{\partial x} = 0.
\label{eq:continuity}
\end{eqnarray}

\begin{figure}[h!]
\includegraphics[width=0.8\textwidth]{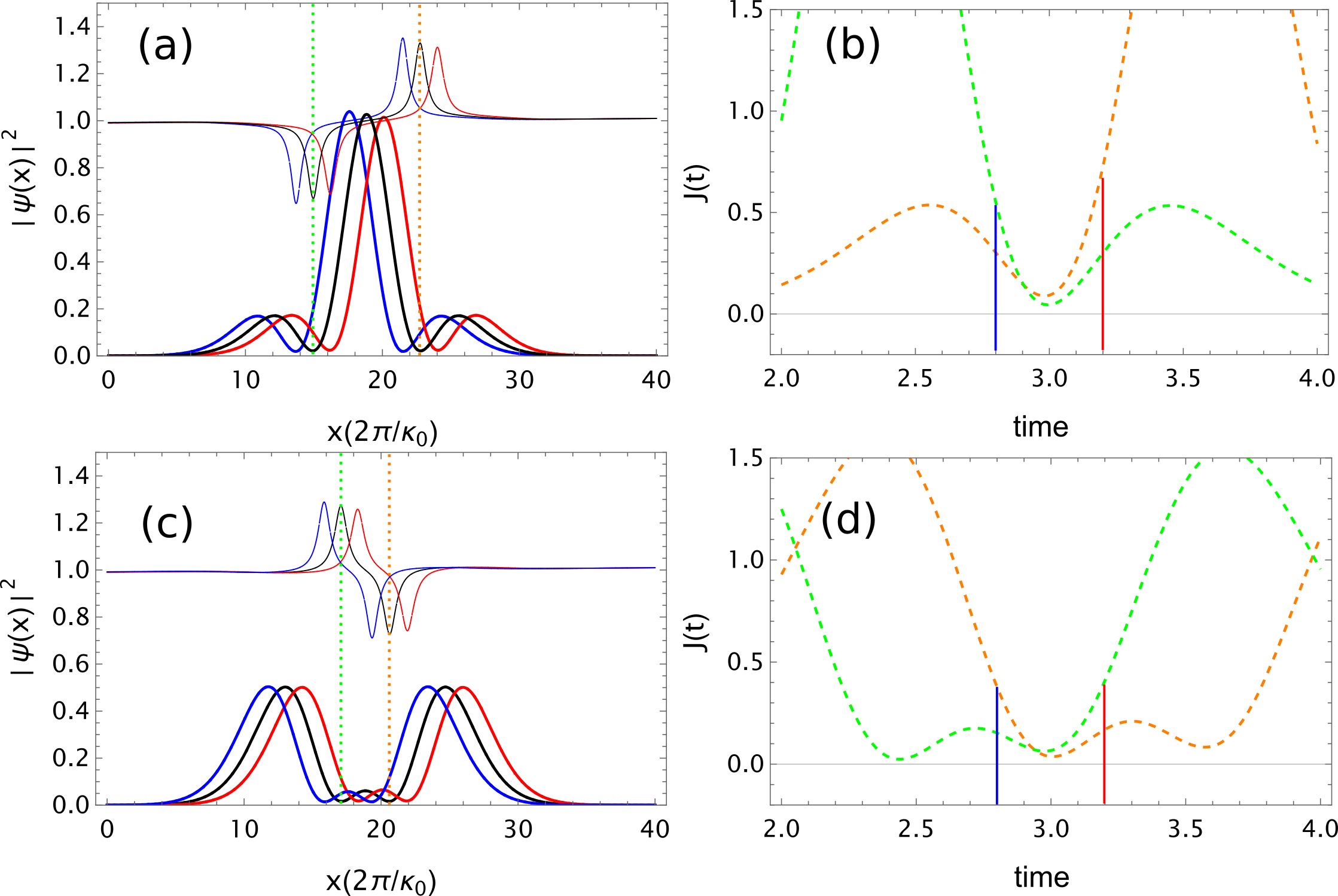}
\caption{ $t=2.8$ in blue line, at $t=3$ in black line and at $t=3.2$ in red line
} \label{fig:Fig4Bis}
\end{figure}

In Figs.~\ref{fig:Fig4Bis}(a) and~\ref{fig:Fig4Bis}(c), we have plotted the probability density for a quantum state with a superbandwidth wavefunction, with the same parameters used in Figs.~\ref{fig:Fig1SM} and~\ref{fig:Fig2SM}, respectively. The different curves in thick solid lines represent the behavior of $|\psi(x,t)|^2$ for a given instant, namely $t=2.8$ (blue line), $t=3$ (black line), and $t=3.2$ (red line), while the corresponding local momenta, $\kappa_l(x,t)$, are shown in thin solid lines with the same colors. In Figs.~\ref{fig:Fig4Bis}(b) and~\ref{fig:Fig4Bis}(d), we show the quantum-mechanical current, as defined in Eq.~\eqref{eq:current}, at two different positions: 
$x = x_{PL}$ (green dashed line), and $x = x_{PR}$ (orange dashed line). 
These are the positions of the transverse planes that are indicated in Figs.~\ref{fig:Fig4Bis}(a) and~\ref{fig:Fig4Bis}(c), as vertical dashed lines. 
Then, we calculate the probability flux (Eq.~\eqref{eq:flux}) that crosses such transverse planes within the instants $t_i = 2.8$ and $t_f = 3.2$. These times are indicated by the blue ($t=2.8$) and red ($t=3.2$) lines in Figs.~\ref{fig:Fig4Bis}(b) and~\ref{fig:Fig4Bis}(d). 

In summary, based on the considerations mentioned above, the calculated value of the probability flux, $F(t_i,t_f)$, is 0.0699 
at $x_{PL}$, and 0.0951 
at $x_{PR}$, for the quantum state of Fig.~\ref{fig:Fig4Bis}(a), while its value is 0.0571 
at $x_{PL}$, and 0.0484 at $x_{PR}$, for the quantum state of Fig.~\ref{fig:Fig4Bis}(c). This indicates that the probability of finding the particle between the planes $x_{PL}$ and $x_{PR}$, i.e., $P(t) = \int_{x_{PL}}^{x_{PR}} dx |\psi(x,t)|^2$, evolves differently with $t$ in each case. Specifically, from Eq.~\eqref{eq:flux}, the difference between the probability fluxes across these two planes, $\Delta(x_{PL}, x_{PR}) \equiv F_{x_{PL}}(t_i, t_f) - F_{x_{PR}}(t_i, t_f)$, can be expressed as
\begin{eqnarray}
\Delta(x_{PL},x_{PR}) &=& \int_{x_{PL}}^{x_{PR}} dx \;|\psi(x,t_f)|^2-\int_{x_{PL}}^{x_{PR}} dx \;|\psi(x,t_i)|^2\nonumber\\
 &=& P(t_f)-P(t_i),
 \label{eq:deltaflux}
\end{eqnarray}
and hence, 
\begin{eqnarray}\label{ eq:deltaflux}
P(t_f) \; \gtrless \; P(t_i) \;\;\; \mathrm{if} \;\;\; \Delta(x_{PL},x_{PR}) \; \gtrless \;0\;. 
\end{eqnarray}

Consequently, the inequality in Eq.~\eqref{eq:deltaflux} tells us that, for a superbandwidth wavepacket with $\alpha=1$ (Fig.\ref{fig:Fig4Bis}(a)), the probability of finding the particle between the positions $x_{PL}$ and $x_{PR}$ decreases with time, given that $F_{x_{PL}}(t_i, t_f) = 0.0699 < 0.0951 = F_{x_{PR}}(t_i, t_f)$. The opposite behavior is observed when $\alpha=1.8$ (Fig.~\ref{fig:Fig4Bis}(c)), since $F_{x_{PR}}(t_i, t_f) = 0.0484 < 0.0571 = F_{x_{PL}}(t_i, t_f)$. In the first case, where $P(t_i) > P(t_f)$, the result seems intuitive: the superbandwidth wavepacket delocalizes over time, as in the case of a freely propagating Gaussian wavepacket—perhaps the best example to understand the time-dependent Schrödinger equation, since it allows for closed analytical solutions. In contrast, in the second case, where $P(t_f) > P(t_i)$, the superbandwidth wavepacket becomes localized as it evolves freely over time, which is, a priori, a counterintuitive behavior.

In fact, when a Gaussian wavepacket $\psi^G(x, t) = \mathcal{F}\left\{e^{-i\frac{\hbar k^2}{2m}t}\phi_0^G(k)\right\}~\left(\phi_0^G(\kappa) \propto e^{-\left(\frac{\kappa-\kappa_0}{\Delta \kappa}\right)^2}\right)$ is considered as the wavefunction of a free non-relativistic quantum particle, it spreads out for $t > 0$. Since this Gaussian wavepacket is the superposition of plane waves, each corresponding to a definite wavevector and having the same phase at $t=0$, the spreading occurs because the different components of the wavepacket acquire different amounts of phase (proportional to $\kappa^2 t$) as they propagate according to the Schrödinger equation and become increasingly out of phase with time.

So, in such a case, the probability $P(t)$ of finding the particle in a finite spatial region of fixed length around the ``center of mass," i.e., around the mean position of the particle $\langle \hat{x} \rangle$, always decreases over time for $t > 0$, according to the probability density given by~\cite{shankarBook}.
\begin{eqnarray}
 |\psi^G(x,t)|^2=\frac{1}{\pi^{1/2}\left( \Delta x^2 + \hbar^2t^2/m^2 \Delta x^2\right)^{1/2}}\exp\left\{-\frac{\left[ x - (\hbar \kappa_0/m) t\right]^2}{ \Delta x^2 + \hbar^2t^2/m^2 \Delta x^2 }\right\},\label{density-gauss}
\end{eqnarray}
being $\Delta x = \Delta \kappa^{-1}$. Furthermore, due to the dependence of the probability density on $t^2$, it is clear that when considering forward propagation from negative values of time, this wavepacket becomes increasingly narrower until $t=0$, where it reaches maximal localization.

Localization effects can, however, be observed during the free propagation of a Gaussian wavepacket if, at $t=0$, the plane wave components have their relative phases arranged such that the Gaussian state acquires a ``quadratic spectral phase" with a sign opposite to that introduced by the propagator in Eq.~\eqref{eq:e5}.
Specifically, this state is described by $\phi^G_{\gamma}(\kappa, t=0) \propto e^{i\frac{\hbar k^2}{2m}\gamma} e^{-\left(\frac{\kappa-\kappa_0}{\Delta \kappa}\right)^2}$, where $\gamma$ is a parameter with values in $\mathbb{R}^+$. The wavepacket corresponding to a free non-relativistic quantum particle with such a momentum distribution is given by $\psi^G_{\gamma}(x, t) = \mathcal{F}\left\{e^{-i\frac{\hbar k^2}{2m}(t - \gamma)} \phi_0^G(\kappa)\right\}$, and its propagation in time can be understood by replacing $t$ with $t - \gamma$ in $\psi^G(x, t)$.  Thus, the free evolution of a quantum state with $\psi^G_{\gamma}(x, t)$ as its wavefunction becomes localized as it evolves in time, until $t=\gamma$, the moment when the system reaches the minimum of the uncertainty relation, $\Delta x \Delta p = \hbar/2$. After this point, the wavepacket spreads like a standard Gaussian state.

In terms of the local momentum $\kappa_l(x, t)$, we can say that the localization (or delocalization) of a wavepacket during free propagation appears to be related to the position $x$ associated with that momentum. Indeed, in Fig.~\ref{fig:Fig1SM}, where the superbandwidth wavepacket delocalizes over time ($P(t_f) < P(t_i)$), the highest local momentum, centered at the value $\kappa_{max}$, leads the wavepacket, while the lowest local momentum, centered at $\kappa_{min}$, lags behind (thin red lines in Figs.~\ref{fig:Fig1SM}(a)-(d)). This behavior is analogous to what is observed in the case of a Gaussian state for times $t > \gamma$, where the local momenta lie on a straight line with a positive slope (thin black lines in Figs.~\ref{fig:Fig1SM}(a)-(d)). In contrast, when the superbandwidth wavepacket becomes localized over time ($P(t_f) > P(t_i)$), the positions of $\kappa_{min}$ and $\kappa_{max}$ are inverted relative to the previous case. This behavior is comparable to that observed for a Gaussian state for times $t < \gamma$, where the higher local momenta are positioned at the back of the wavepacket.


 \subsection{Classical analysis and \textit{interference time}}

In this section, we analyze the free propagation of an ensemble of classical particles. This analysis provides, through comparison with the propagation of a Gaussian wavepacket, a schematic perspective on the phenomena previously presented in this work. Additionally, we define a characteristic time, referred to as the \textit{interference time}, to estimate the moment when the local oscillations, centered at the values $\kappa_{min}$ and $\kappa_{max}$, vanish. It is important to note that this classical analysis should not be interpreted as a semiclassical approach.

Let us first consider a set of $N$ particles with mass $m$ and momentum $p_i (i=1,2,\dots , N)$ moving in the positive $x$ direction. If, at a given time $t = t_0$, all the particles are at the same position, $x_i(t_0) = x_0$, then at a later time $t = t_p$, the position of each particle will depend on its momentum, such that $x_i(t_p) < x_j(t_p)$ if $p_i < p_j$. This behavior is analogous to what is observed in the free propagation of a Gaussian wavepacket, where the higher local momenta (see the thin black line in Figs.~\ref{fig:Fig1SM}(a)--(d)) are localized at the front of the wavepacket, as the slope of $\kappa_l(x, t)$ at a fixed time is positive. 

\begin{figure}[h!]
\includegraphics[width=0.8\textwidth]{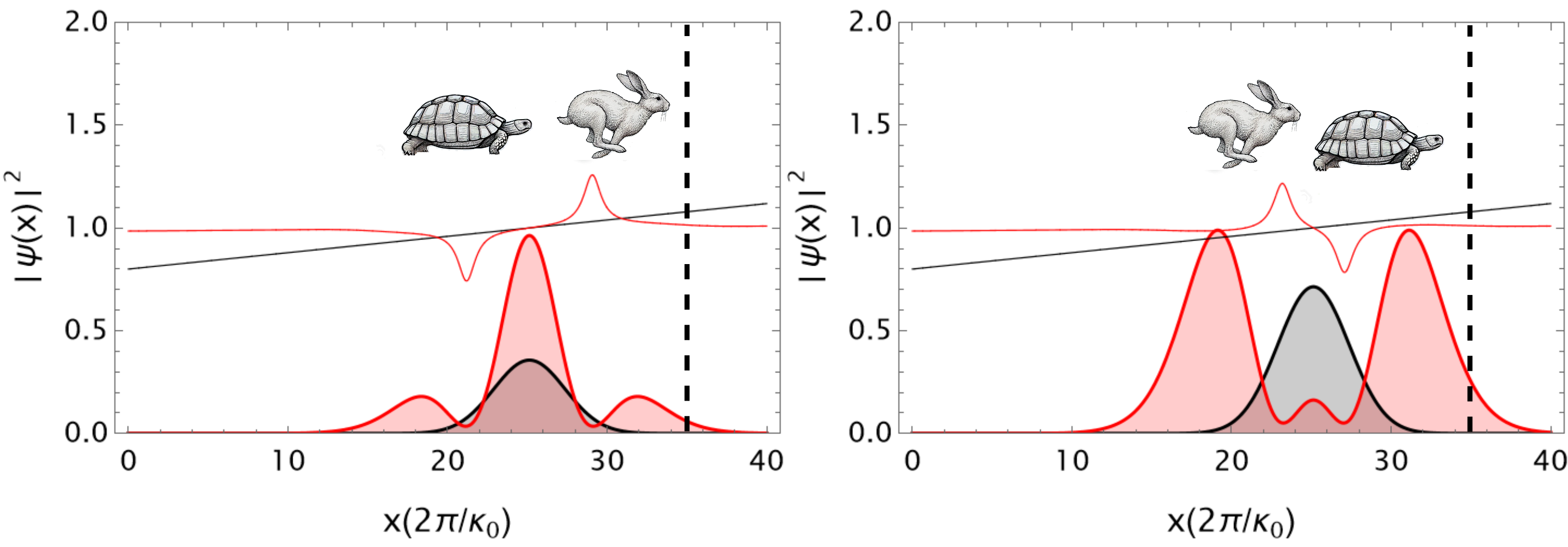}
\caption{Representation of the tale of the hare and the tortoise: In the left panel, the hare arrives before the tortoise, while in the right panel, the tortoise arrives before the hare. 
} \label{fig:Fig-T-L}
\end{figure}


However, it can be observed that the free propagation of a superbandwidth wavepacket exhibits two regimes of propagation. In the first regime, when $\alpha < \alpha_C$, the local momentum $\kappa_{max}$ (\textit{hare}) is ahead of $\kappa_{min}$ (\textit{tortoise}) (see Fig.~\ref{fig:Fig-T-L}, left). At a hypothetical finish line (dashed line), the \textit{hare} reaches the finish before the \textit{tortoise}. In the second regime, the local momentum $\kappa_{min}$ (\textit{tortoise}) is ahead of $\kappa_{max}$ (\textit{hare}), and the \textit{tortoise} reaches the finish before the \textit{hare} (see Fig.~\ref{fig:Fig-T-L}, right).

In Ref.~\cite{Neyra2023}, we also demonstrated that the appearance of these local oscillations, which lie beyond the Fourier spectrum, occurs when the pulse propagates through a dispersive medium over a distance nearing the point at which the pulse's width doubles. Beyond this distance, the temporal Fourier transform is obtained, and these local oscillations vanish, as discussed in Ref.~\cite{goda2013dispersive}. This observation implies that the phenomena described are present for times $t$ during which the wave function spreads by a factor of two or less. The spreading of a Gaussian wavepacket, $\Delta x_t$, can be expressed in terms of the spatial standard deviation $\Delta x$ or the momentum standard deviation $\Delta p$, where $\Delta p = \hbar \Delta \kappa$, as
\begin{equation}
 \Delta x_t^2=\Delta x^2\left(1 + \left(\frac{\hbar t}{2m\Delta x^2}\right)^2\right).
\label{eq:deltax}
\end{equation}
If $\Delta x\Delta p=\hbar/2 $, the last expression results
\begin{equation}
 \Delta x_t^2=\Delta x^2\left(1 + \left(\frac{2\Delta p^2 t}{m\hbar}\right)^2\right).
\label{eq:deltax_bis}
\end{equation}
For example, if the spreading satisfies the condition $\Delta x_t^2 = 2\Delta x^2$, we obtain the relation $\frac{2\Delta p^2 t}{m\hbar} = 1$. The time $t$ will therefore be of the order of $\frac{m\hbar}{2\Delta p^2}$ or, equivalently, $t = \frac{2m\Delta x^2}{\hbar}$.

\begin{figure}[h!]
\includegraphics[width=0.8\textwidth]{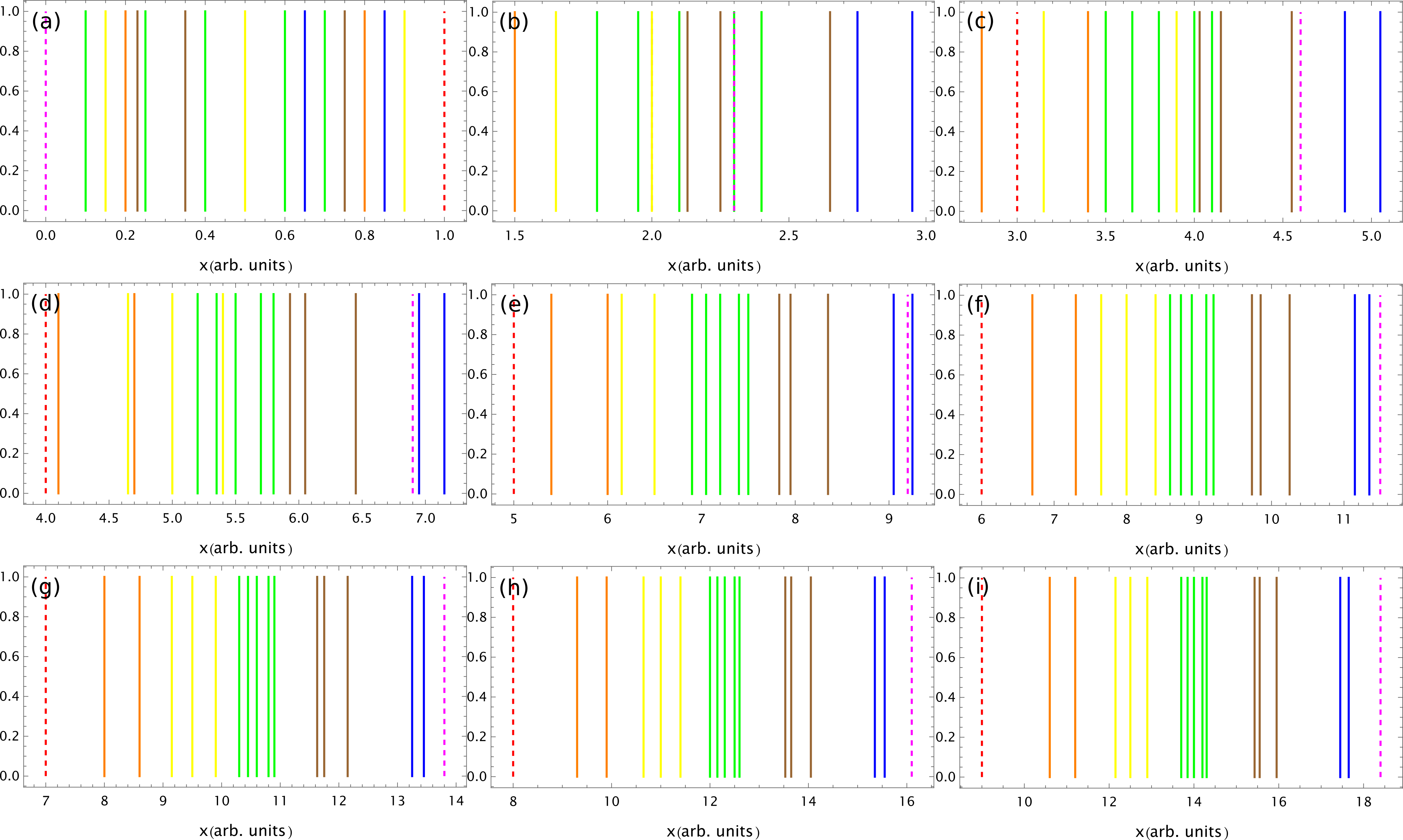}
\caption{Schematics representation of a classical set of particles moving freely with different velocities. The vertical color lines indicate a particle and its color its velocity (see the main text for detail). In panel (a) is shown the state of the particles at $t=0$ with a random distribution in their initial conditions. The other panels indicate a different time evolution, where (b) represent $t=1$, (c) $t=2$, (d) $t=3$, (e) $t=4$, (f) $t=5$, (g) $t=6$, (h) $t=7$ and (i) $t=8$.  
} \label{fig:Fig-par}
\end{figure}

Now, consider a classical scenario in which we have a set of $N$ particles of mass $m$ moving freely. At $t = 0$, these particles are located within the spatial interval $[x_l, x_h]$ with a random distribution of momentum in the range $[p_l, p_h]$. At a later time $t = t_p$, two of these particles, with momenta $p_i < p_j$, will be at positions $x_i(t_p)$ and $x_j(t_p)$, respectively. Here, we cannot guarantee that the particle with the higher momentum satisfies $x_i(t_p) < x_j(t_p)$, because at $t = 0$, it is possible that $x_j(0) < x_i(0)$. However, we can assert that there exists a sufficiently large time $t_l$ such that for all times $t > t_l$, $x_i(t) < x_j(t)$.

An estimated order of magnitude for the time $t_l$, which we will call the \textit{interference time} $t_I$, can be obtained when the particle with the highest momentum $p_h$ is at position $x_l$, and the particle with the lowest momentum $p_l$ is at $x_h$ at $t = 0$. This time can be easily determined using the condition $x_i(t_I) = x_j(t_I)$, which implies $t_I = \frac{m\delta x}{\delta p}$, where $\delta x = x_h - x_l$ and $\delta p = p_h - p_l$. Furthermore, if we impose the condition $\delta x \delta p = \hbar$, we find the value of $t_I$ to be $t_I = \frac{m\delta x^2}{\hbar}$. It can be seen that this time is equivalent to the quantum analysis discussed in the previous paragraph, i.e., $t_I = \frac{m\delta x^2}{\hbar} \approx \frac{2m\Delta x^2}{\hbar}$.

In a quantum mechanical scenario involving the free evolution of a wavepacket, the time $t_I = \frac{m\Delta x^2}{\hbar} \ll t$ provides an estimate for when the interference ``inside" the wavepacket disappears. At this point, the momentum distribution within the wavepacket becomes well-separated spatially, corresponding to the momentum distribution given by $|\phi_0(\kappa)|$. In other words, the shape of the wavefunction is $|\psi(x,t)|^2 \propto |\phi_0(k)|^2$, as discussed in Ref.~\cite{goda2013dispersive}.

The above analysis can also be understood classically by considering the difference between two consecutive trajectories, $x_i(t) = \frac{p_i t}{m} + x_i(0)$ and $x_{i+1}(t) = \frac{p_{i+1} t}{m} + x_{i+1}(0)$. The difference in their positions is given by $x_{i+1}(t) - x_i(t) = \Delta x_{i+1}(t) = \frac{\Delta p_{i+1} t}{m} + \Delta x_{i+1}(0)$. If the time is sufficiently long, such that $\Delta x_{i+1}(0) \ll \frac{\Delta p_{i+1} t}{m}$, then $\Delta x_{i+1}(t) \propto \Delta p_{i+1}$. This shows that once the particles are far enough from the initial spatial interval $\Delta x_{i+1}(0)$, the difference in their positions becomes directly proportional to the difference in their momenta.

To visualize the above discussion, Fig.~\ref{fig:Fig-par} presents a set of identical classical particles moving freely in the $x$-direction. The particles are represented by vertical colored lines, where the color indicates the velocity of each particle. Using units of $[x] = [t] = 1$, the particle velocities are as follows: red dashed line ($v = 1$), orange line ($v = 1.3$), yellow line ($v = 1.5$), green line ($v = 1.7$), brown line ($v = 1.9$), blue line ($v = 2.1$), and magenta dashed line ($v = 2.3$). Randomly selected spatial initial conditions were assigned to each particle, with the constraint that the slowest particle (red dashed line) is initially at $x_h$, and the fastest particle (magenta dashed line) is initially at $x_l$. The figure consists of nine subplots, each illustrating the system's evolution at a different time. Specifically, Fig.~\ref{fig:Fig-par}(a) shows the initial spatial distribution at $t = 0$, Fig.~\ref{fig:Fig-par}(b) the evolution at $t = 1$, Fig.\ref{fig:Fig-par}(c) at $t = 2$, Fig.~\ref{fig:Fig-par}(d) at $t = 3$, Fig.\ref{fig:Fig-par}(e) at $t = 4$, Fig.\ref{fig:Fig-par}(f) at $t = 5$, Fig.\ref{fig:Fig-par}(g) at $t = 6$, Fig.\ref{fig:Fig-par}(h) at $t = 7$, and Fig.~\ref{fig:Fig-par}(i) at $t = 8$.

From these figures, we can observe how the particles, as they evolve over time, ``arrange" themselves such that those with larger momenta move ahead of those with smaller momenta. In this case, the interference time is $t_I = \frac{m\delta x^2}{\hbar} = 1$, assuming the mass $m = 1$. This corresponds to the configuration shown in Fig.~\ref{fig:Fig-par}(b). For times on the order of $t_I$, the particles begin to cross each other, which can be seen as an analogy to the interference of different Fourier components within a wavepacket. For times $t_I \ll t$, however, the spatial distribution of the particles is primarily determined by their momentum distribution.


In Ref.~\cite{trillo2023quantum}, an equivalent parameter was introduced to determine the region where some quantum advantage may exist for transport tasks. This parameter can be obtained from the time $t_I$ defined here as $\frac{m\Delta x^2}{\hbar t_I}$. In that work, it is shown that when $\frac{m\Delta x^2}{\hbar t_I} \ll 1$, no quantum advantage is present. This condition corresponds to the limit where $\frac{m\Delta x^2}{\hbar} \ll t$.


 \subsection{Bohmian mechanical analysis}

In this section, we analyze the evolution of the superbandwidth wavepacket through the lens of Bohmian mechanics~\cite{Bohm1952}, an alternative formulation of quantum mechanics. In this framework, the temporal evolution of a quantum system is mathematically described by a guiding equation, which allows quantum mechanics to be interpreted in terms of particle trajectories and provides a deterministic perspective on the theory. Unlike Newton’s equations, the guiding equation incorporates a quantum potential that depends on the wavefunction as defined by the Schrödinger equation. This quantum potential introduces non-local effects, meaning that a particle's trajectory is influenced by the entire configuration of trajectories encoded in the wavefunction.

For a single particle, the complex wave function $\psi(x,t)$ of a quantum state can be written in polar form as
\begin{equation}
 \psi(x,t)= R(x,t) e^{i S(x,t)/\hbar} ~.
\label{eq:Polar}
\end{equation}
%
By substituting the expression in Eq.~\eqref{eq:Polar} into the Schrödinger equation (Eq.~\eqref{eq_Schrodinger}), the continuity equation and the Hamilton-Jacobi equation can be derived~\cite{sanz2008trajectory}. From these, the one-dimensional guiding equation is obtained, expressed as:
\begin{equation}
 v_B(t)=\frac{d x_B(t)}{dt}=\frac{1}{m} \frac{\partial S(x_B(t),t)}{\partial x} ~.
\label{eq:guiding equation}
\end{equation}

Here, an infinite set of trajectories $x_B(t)$ depends on the initial conditions $x_{B0} = x_B(t = 0)$, which are randomly selected under the constraint that they lie within the spatial interval defined by $R(x, 0)$.

Equation~\eqref{eq:guiding equation} establishes a mathematical equivalence with Eq.~\eqref{eq:local_momentum_b}, expressed as $m v_B(t) \equiv \kappa_l(x,t)$. This indicates that it is possible to associate the velocity of the Bohmian trajectories with the specific local momenta centered at $\kappa_{min}$ and $\kappa_{max}$.

\begin{figure*}[h!]
\includegraphics[width=1\textwidth]{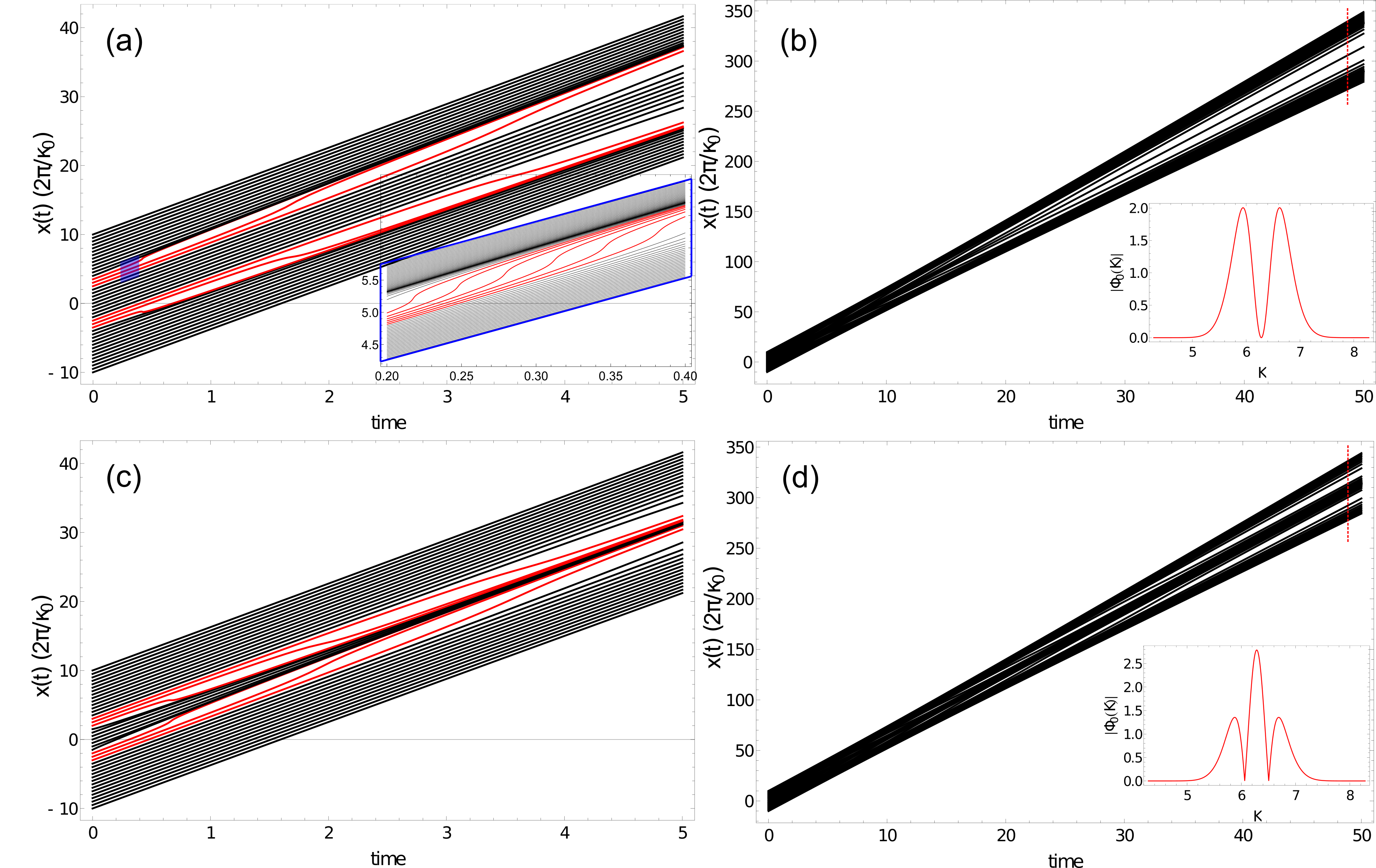}
\caption{Bohmian trajectories of the free propagation of a superbandwidth wavepacket. Panel (a) shows a superbandwidth wavepacket with $\alpha=1$ propagating from $t=0$ to $t=5$. The trajectories in red line are associated with the the special local momenta centered at $\kappa_{max}$ and $\kappa_{min}$. The insets show a zoom in the region determined by the blue rhomboid. Panel (c) shows a superbandwidth wavepacket with $\alpha=1.8$ propagating from $t=0$ to $t=5$. The panels (b) and (d) show the superbandwidth wavepackets showing in panel (a) and (c) propagating from $t=0$ to $t=50$. The insets in red line represent the momenta distribution $|\phi_0(k)|$. 
} \label{fig:Fig-Bohm}
\end{figure*}

In Figs.~\ref{fig:Fig-Bohm}(a) and \ref{fig:Fig-Bohm}(c), the Bohmian trajectories calculated for the superbandwidth wavepackets shown in Figs.~\ref{fig:Fig1SM} and \ref{fig:Fig2SM} are depicted, corresponding to $\alpha = 1$ (Fig.\ref{fig:Fig-Bohm}(a)) and $\alpha = 1.8$ (Fig.~\ref{fig:Fig-Bohm}(c)). We present a set of 40 trajectories with initial conditions in the spatial interval $[-10,10]$, showing their temporal evolution from $t = 0$ to $t = 5$. The trajectories associated with the specific local momenta, centered at $\kappa_{min}$ and $\kappa_{max}$, are represented by red lines, while the other trajectories are shown in black.

From Figs.~\ref{fig:Fig-Bohm}(a) and \ref{fig:Fig-Bohm}(c), it can be observed that the behavior of these special trajectories, represented by red lines, deviates significantly from a straight line. In Fig.~\ref{fig:Fig-Bohm}(a), at the front of the wavepacket, the trajectories accelerate, reaching a high velocity (indicated by the steep slope of the curve) that exceeds the Fourier ``velocity" spectrum of the wavepacket. This steep slope gradually diminishes as the wavepacket evolves over time. A similar behavior is observed in Fig.~\ref{fig:Fig1SM}, where $\kappa_{max}$ begins to vanish. At the back of the wavepacket, the behavior is opposite: the trajectories decelerate, the slope is smaller than that of the black trajectories, and the behavior corresponds to $\kappa_{min}$ in Fig.~\ref{fig:Fig1SM}. For better visualization of the behavior of the special trajectories, we plot, in Fig.~\ref{fig:Fig-Bohm}(a), a zoomed view of the spatial region $[4,6]$ between times $t = 0.2$ and $t = 0.4$ (see the blue rectangular region). This allows for a more detailed observation of the evolution of the special trajectories.

Figure~\ref{fig:Fig-Bohm}(c) shows the inverse situation compared to Fig.~\ref{fig:Fig-Bohm}(a). Here, the trajectories with faster velocities are located at the back of the wavepacket, while those with slower velocities are at the front. This result aligns with Fig.~\ref{fig:Fig2SM}, where the local momenta $\kappa_{max}$ and $\kappa_{min}$ are inverted with respect to Fig.~\ref{fig:Fig1SM}. Another noteworthy feature observed in the propagation of the superbandwidth wavepacket through Bohmian analysis is the alternation between localization and delocalization of the superbandwidth wavepackets, as discussed at the end of Section A. Figure \ref{fig:Fig-Bohm}(a) illustrates how the density of trajectories in the central region (bounded by the  trajectories in red) decreases as the superbandwidth wavepackets evolve over time. This decrease in trajectory density occurs faster than in the case of a Gaussian wavepacket (see Eq.~\eqref{eq:wave_packet_momentum} with $\alpha = 0$), meaning that the spreading of a superbandwidth wavepacket is faster than that of a Gaussian wavepacket. Conversely, Fig.~\ref{fig:Fig-Bohm}(c) demonstrates the localization of the superbandwidth wavepackets in the central region as they evolve over time.

Finally, in Figs.~\ref{fig:Fig-Bohm}(b) and~\ref{fig:Fig-Bohm}(d), the propagation of the superbandwidth wavepackets shown in Figs.~\ref{fig:Fig-Bohm}(a) and \ref{fig:Fig-Bohm}(c), respectively, is depicted over a time interval from $t = 0$ to $t = 50$, far beyond $t_I$ (see Section B). Here, it can be observed that the trajectories take the form of the momentum spectrum (see dashed red line). The momentum spectra $|\phi_0(k)|$ for each case are shown in the insets with red lines. Here, for times $t_I \ll t$, the trajectories become straight lines, maintaining constant velocity with no changes in their state. In this regime, the effect of the quantum potential as a ``force" vanishes.

Another perspective on the time $t_I$ is provided in Ref.~\cite{sanz2008trajectory}, which analyzes the free Gaussian propagation in terms of an effective time $\tau$ and defines velocities related to the spreading of the wavepacket. The behavior of these special trajectories, i.e., those that deviate sharply from a straight line—is analogous to the trajectories observed in a double-slit experiment (Ref.~\cite{deotto1998bohmian}). Similarly, Ref.~\cite{braverman2013proposal} reports comparable results, where an entangled pair of photons in a double-slit setup exhibits superluminal characteristics.

\section{Conclusions}

In conclusion, we have studied the dynamics of the free propagation of a superbandwidth wavepacket using the time-dependent Schrödinger equation, providing deeper insights into its distinctive behavior within the framework of Bohmian mechanics.

As expected, due to the mathematical similarity between laser pulse propagation in a dispersive medium and the free propagation of a wavepacket, local momenta emerge during the propagation of a superbandwidth wavepacket, exhibiting both sub- and super-oscillatory characteristics. Moreover, these special local momenta can be modulated over time by varying the parameter $\alpha$, such that lower momenta appear to propagate faster than higher momenta.

Additionally, we demonstrated that the superbandwidth wavepacket can either localize or delocalize during propagation, depending on the value of the parameter $\alpha$. Specifically, the delocalization or spreading of the superbandwidth wavepacket occurs at a faster rate compared to a Gaussian wavepacket (see Eq.\eqref{eq:wave_packet_momentum} for $\alpha = 0$), where the spreading of the Gaussian wavepacket is governed by Eq.~\eqref{eq:deltax}. Consistent with the concept of "local phenomena," we refer to this behavior as \textit{super-delocalization} or \textit{super-spreading}. Conversely, the localization or compression of the superbandwidth wavepacket is more challenging to interpret within the framework of quantum mechanics.

Finally, we conclude that the phenomena described above occur for times on the order of the characteristic time $t_I$, which we define as the "interference time." Beyond this regime, when $t_I \ll t$, all of these interferometric effects vanish.



\section*{Acknowledgments}

The present work is supported by the National Key Research and Development Program of China (Grant No.~2023YFA1407100), the Guangdong Province Science and Technology Major Project (Future functional materials under extreme conditions - 2021B0301030005) and the Guangdong Natural Science Foundation (General Program project No. 2023A1515010871). E. G. N. and L. R.~acknowledge to Consejo Nacional de Investigaciones Cient\'ificas y T\'ecnicas (CONICET).

\bibliography{main}

\end{document}